\documentstyle[preprint,prc,aps]{revtex}
\begin{document}
\draft
\title{The \bbox{E2} contribution to the \bbox{{}^8{\rm B}\rightarrow
p + {}^7{\rm Be}} Coulomb dissociation cross section}
\author{K. Langanke and T. D. Shoppa}
\address{W. K. Kellogg Radiation Laboratory, 106-38\\
California Institute of Technology, Pasadena, California 91125 USA}
\date{\today}
\maketitle

\begin{abstract}
We have calculated the $E1$ and $E2$ contributions to the low-energy
${}^8{\rm B}+ {}^{208}{\rm Pb} \rightarrow p+{}^7{\rm Be}
+{}^{208}{\rm Pb}$ Coulomb dissociation cross sections using the
kinematics of a recent experiment at RIKEN. Using a potential model
description of the ${}^7{\rm Be} (p,\gamma) {}^8{\rm B}$ reaction, we
find that the $E2$ contributions cannot {\it a~priori} be ignored in
the analysis of the data. Its inclusion reduces the extracted
${}^7{\rm Be} (p,\gamma) {}^8{\rm B}$ $S$-factor at solar energies by
about 25\%.
\end{abstract}

\narrowtext
The ${}^7{\rm Be} (p,\gamma) {}^8{\rm B}$ reaction plays a crucial
role in the solar neutrino puzzle, as its rate is directly
proportional to the flux of those high-energy neutrinos to which the
${}^{37}$Cl and Kamiokande detectors are particularly sensitive
\cite{Bahcall}. While the energy dependence of the low-energy
${}^7{\rm Be} (p,\gamma) {}^8{\rm B}$ cross section is believed to be
sufficiently well known \cite{Johnson}, the absolute cross section at
solar energies ($E\approx20$~keV) is rather uncertain as the two
measurements of the cross section that extend lowest in energy
disagree by about 25\% in magnitude \cite{Kavanagh,Filippone}. The
recent availability of radioactive beam facilities offers the
possibility of resolving this discrepancy indirectly by measuring the
Coulomb dissociation of a ${}^8$B nucleus in the field of a
heavy-target nucleus like ${}^{208}$Pb. Performing such an experiment
at carefully chosen kinematics to minimize nuclear-interaction
effects and assuming the break-up as a one-step process in which a
single virtual photon is absorbed, the Coulomb dissociation is the
inverse of the radiative capture process \cite{Baur}.

Recently an experiment at RIKEN measured the ${}^8{\rm B}+
{}^{208}{\rm Pb} \rightarrow p+{}^7{\rm Be} +{}^{208}{\rm Pb}$
dissociation cross section at the high incident energy of 46.5~MeV/u
\cite{Gai}. Using the semi-classical formulas of Ref.~\cite{Baur},
the Coulomb dissociation cross section was translated into
$S$-factors for the ${}^7{\rm Be} (p,\gamma) {}^8{\rm B}$ radiative
capture process. From this it was concluded that the ${}^7{\rm Be}
(p,\gamma) {}^8{\rm B}$ $S$-factor at solar energies is likely to be
smaller than 20~eV-b, supporting the lower \cite{Filippone} of the
two direct ${}^7{\rm Be} (p,\gamma) {}^8{\rm B}$ measurements.

In Ref.~\cite{Gai} the Coulomb dissociation was analyzed as a pure
$E1$ break-up process, ignoring possible $E2$ contributions. This
assumption is certainly valid for the radiative capture reaction, in
which the $E1$ cross section is estimated to dominate $E2$ captures
by nearly 3 orders of magnitude at low energies \cite{Kim}. However,
as the number of virtual photons strongly favors $E2$ transitions,
the ratio of $E2$-to-$E1$ Coulomb dissociation cross sections
($\sigma^{cd}_{E2}/ \sigma^{cd}_{E1}$) is significantly different,
relatively enhancing the importance of $E2$ transitions. As has been
shown in studies of the ${}^6{\rm Li} +{}^{208}{\rm Pb} \rightarrow
D+\alpha +{}^{208}{\rm Pb}$ \cite{Typel}, ${}^7{\rm Li} +{}^{208}{\rm
Pb} \rightarrow t+\alpha +{}^{208}{\rm Pb}$ \cite{Typel1}, and
${}^{16}{\rm O} +{}^{208}{\rm Pb} \rightarrow \alpha+{}^{12}{\rm C}
+{}^{208}{\rm Pb}$ \cite{Shoppa} reactions, this enhancement can
amount to more than two orders of magnitude, depending on the
kinematics of the break-up process.

In the following we will estimate the $E2$ contribution to the
${}^8{\rm B}+ {}^{208}{\rm Pb} \rightarrow p+{}^7{\rm Be}
+{}^{208}{\rm Pb}$ cross section at the kinematics used in the RIKEN
experiment. As in the analysis of Ref.~\cite{Gai} we will use the
semi-classical formalism of Baur {\it et~al.} \cite{Baur} to connect
the break-up cross section to the radiative capture cross section. We
adopt the $E1$ and $E2$ capture cross sections from the ${}^7{\rm Be}
(p,\gamma) {}^8{\rm B}$ potential model calculation of Kim {\it et
al.} \cite{Kim}, which has also served as a theoretical guideline in
Ref.~\cite{Gai}.

The RIKEN experiment \cite{Gai} has measured the double differential
cross section for the ${}^8{\rm B}+ {}^{208}{\rm Pb} \rightarrow
p+{}^7{\rm Be} +{}^{208}{\rm Pb}$ Coulomb dissociation reaction as a
function of the Rutherford scattering angle $\theta_R$ and the
center-of-mass energy in the $p+{}^7{\rm Be}$ system, $E_{17}$. One
has \cite{Baur}
\widetext
\begin{equation}
\frac{d^2\sigma}{d\Omega_R dE_{17}}=
\sum_{J_f,\lambda}
\left({\frac{Z_{Pb} e}{\hbar v_i}}\right)^2
a^{-2\lambda+2} B(E\lambda, J_i\rightarrow J_f, E_{17})
\frac{df_{E \lambda} (\theta_R , \xi)}{d \Omega_R}\;,
\end{equation}
\narrowtext
where
\begin{equation}
a = \frac{Z_{B} Z_{Pb} e^2}{\mu v_i v_f}
\end{equation}
is the half-distance of closest approach and
\begin{equation}
\xi = \frac{Z_{B} Z_{Pb} e^2}{\hbar}
(\frac{1}{v_f} - \frac{1}{v_i})\;.
\end{equation}
is the adiabaticity parameter. Here, $v_i, v_f$ denote the relative
velocities between projectile and target in the initial and final
channels, while $Z_k$ is the atomic number of the fragment $k$. The
reduced mass $\mu$ is defined between the ${}^8$B and the
${}^{208}$Pb nuclei. The quantity $\frac{df_{E \lambda} (\theta_R,
\xi)} {d \Omega_R}$ can be calculated in the straight-line
approximation from the formulae given in Ref.~\cite{Alder}. Finally,
the $B(E\lambda)$ matrix elements are related to the respective
partial ${}^7{\rm Be} (p,\gamma) {}^8{\rm B}$ cross sections via
\[
\sigma_{E \lambda}^{J_f \rightarrow J_i} (p+{}^7{\rm Be}
\rightarrow{}^8{\rm B}+\gamma)=\]
\begin{equation}
\frac{16 \pi^3 (\lambda +1)}{\lambda [(2\lambda +1)!!]^2}
\left({\frac{E_{\gamma}}{\hbar c}} \right)^{2 \lambda +1}
\frac{\hbar^2}{2\mu_{17} E_{17}}
B(E \lambda, J_i \rightarrow J_f, E_{17} )\;,
\end{equation}
where $J_i, J_f$ are the total angular momenta of the initial and
final states in the Coulomb dissociation reaction, $\mu_{17}$ is the
reduced mass of the $p+{}^7{\rm Be}$ system and $E_{\gamma}$ denotes
the photon energy. We have calculated the $B(E\lambda)$ matrix
elements from the partial ${}^7{\rm Be} (p,\gamma) {}^8{\rm B}$ $E1$
and $E2$ cross sections as given in Ref.~\cite{Kim}. This $E1$ cross
section agrees well with the measured ${}^7{\rm Be} (p,\gamma)
{}^8{\rm B}$ data. Due to the lack of better experimental
constraints, the initial scattering states for the $E2$ cross section
have been calculated by using the same $l$-independent radial optical
potential fitted to the $M1$ resonance at 633~keV. It should be noted
that the $E2$ cross section is not tested directly against
experimental data and might thus be viewed as somewhat uncertain.
Nevertheless the potential model estimate given in Ref.~\cite{Kim} is
probably accurate enough to determine whether $E2$ contributions can
be ignored in the ${}^8{\rm B}+ {}^{208}{\rm Pb} \rightarrow
p+{}^7{\rm Be} +{}^{208}{\rm Pb}$ cross sections.

The authors of Ref.~\cite{Gai} have studied the ${}^8{\rm B}+
{}^{208}{\rm Pb} \rightarrow p+{}^7{\rm Be} +{}^{208}{\rm Pb}$
reaction at various relative energies $E_{17}$ between 600~keV and
about 2~MeV and at Rutherford scattering angles $\theta_R \leq
6^{\circ}$. In Fig.~1 we show the ratio of virtual photon numbers for
$E2$ and $E1$ transitions in the ${}^8{\rm B}+ {}^{208}{\rm Pb}
\rightarrow p+{}^7{\rm Be} +{}^{208}{\rm Pb}$ reaction covering the
experimental energy range and at some typical $\theta_R$-values. We
observe that the $E2/E1$ enhancement increases with angles, while it
decreases with relative energy. While the enhancement is smaller than
100 at all experimentally relevant energies at the smallest angles
data have been taken, it already amounts to more than 100 at
$\theta_R =2^{\circ}$ for the astrophysically important energy range
$E_{17}\leq 1$~MeV. Considering that the ratio of partial $E1$ to
$E2$ ${}^7{\rm Be} (p,\gamma) {}^8{\rm B}$ cross sections is
estimated \cite{Kim} to be less than about a factor 1000, we expect
that the $E2$ contribution to the ${}^8{\rm B}+ {}^{208}{\rm Pb}
\rightarrow p+{}^7{\rm Be} +{}^{208}{\rm Pb}$ cross section cannot be
ignored at angles $\theta_R\geq 2^{\circ}$ and energies
$E_{17}\leq1$~MeV. This conjecture is confirmed in Fig.~1 where we
have plotted ($\sigma^{cd}_{E2}/ \sigma^{cd}_{E1}$). The maximum of
this ratio at around $E_{17} = 633$~keV is related to the lowest
$1^+$ resonance in ${}^8$B. The main electromagnetic decay of this
state is by $M1$ transition to the ${}^8$B ground state with
$J^{\pi}=2^+$. While an $E1$ Coulomb excitation of this resonance is
forbidden by parity, an $E2$ excitation is allowed leading to a
particularly large $E2$ contribution around the resonance energy.
With the partial $E1$ and $E2$ cross sections of Ref.~\cite{Kim}, one
finds that the $E2$ process dominates the total ${}^8{\rm B}+
{}^{208}{\rm Pb} \rightarrow p+{}^7{\rm Be} +{}^{208}{\rm Pb}$ cross
section at angles $\theta_R\geq 4^{\circ}$.

Despite possible uncertainties in the potential model calculation,
the $E2$ contribution will contribute significantly to the total
Coulomb break-up cross section in the vicinity of the resonance and
has to be taken into account in the data analysis. A precise
measurement of the Coulomb dissociation cross section at the
resonance energy and at angles $\theta_R>2^{\circ}$ will determine
the strength of the partial $E2$ capture cross section at this energy
and thus place an important constraint on the theoretical modeling of
this cross section. Of course, it would be desirable to measure the
triple-differential Coulomb dissociation cross section $\frac{d^3
\sigma} {d \Omega_R d \Omega_{17} d E_{17}}$, where $\Omega_{17}$
defines the angle between the proton and the ${}^7$Be nucleus out of
the scattering plane. This quantity is sensitive to the interference
of $E1$ and $E2$ Coulomb break-up transitions \cite{Weber}.

In Ref.~\cite{Gai} the ${}^7{\rm Be} (p,\gamma) {}^8{\rm B}$
$S$-factors at different relative energies (binned into intervals of
200~keV width) have been determined by fitting the
double-differential ${}^8{\rm B}+ {}^{208}{\rm Pb} \rightarrow
p+{}^7{\rm Be} +{}^{208}{\rm Pb}$ yields for fixed energy as a
function of Rutherford scattering angle (binned into intervals of
width 1~degree). As mentioned above, only $E1$ Coulomb break-up has
been considered. We will now discuss how significantly $E2$ break-up
might contribute to the data of Ref.~\cite{Gai}. As we do not know
the detector efficiency function, a direct calculation of the yields
is not possible. Assuming that the detector efficiency is the same
for $E1$ and $E2$ contributions, we take the yield curves in Fig.~2
of Ref.~\cite{Gai} and multiply by $(\sigma^{cd}_{E1}+
\sigma^{cd}_{E2})/ \sigma^{cd}_{E1}$. Here we have averaged the cross
sections over the same angular and energy bins as in Ref.~\cite{Gai}.
We find that the ratio is rather robust against this averaging. The
relative importance of the $E2$ contribution can be seen as the
difference between the dashed ($E1+E2$) and dotted ($E1$) curves in
Fig.~2. As expected, $E2$ Coulomb break-up is most important at the
energy interval centered around $E_{17}=0.6$~MeV, which covers the
$1^+$ resonance at 633~keV. Here we find a noticeable change of the
yield curve in both magnitude and shape. At the higher energies, the
effect of the $E2$ break-up is less pronounced than at the resonance
energy leading to no significant change in the yield pattern.

As the $E1$ and $E2$ break-up parts add in the double-differential
cross section (1), the presence of the $E2$ component in the data
will reduce the partial $E1$ cross section compared to the one
deduced in Ref.~\cite{Gai}, which ignored possible $E2$
contributions. We have fitted the data of Ref.~\cite{Gai} to our
($E1+E2$) yield curves by multiplying the calculated yields with a
parameter $\alpha(E_{17})$ which has been determined by
$\chi^2$-minimization. As our yields are normalized to the $E1$
yields of Ref.~\cite{Gai}, the partial $E1$ ${}^7{\rm Be} (p,\gamma)
{}^8{\rm B}$ $S$-factor extracted from the data scales by the same
parameter $\alpha$. We find that at the resonance ($E_{17}=0.6$~MeV)
the data agree noticeably better with our ($E1+E2$) yield curve than
with a pure $E1$ pattern (Fig.~2); the $\chi^2$ between the two fits
is reduced by 30\%. Thus, the experimental data at this energy show
the presence of the $1^+$ resonance. We obtain a best-fit value of
$\alpha(0.6) =0.66\pm0.08$. At the two other energies our fit
procedure results in $\alpha(0.8)= 0.82\pm0.16$ and $\alpha(1.0)
=0.77\pm0.17$, while the $\chi^2$-values are about the same for pure
$E1$ and our ($E1+E2$) yields pattern. The values of the parameter
$\alpha(E_{17})$ translates into the partial $E1$ ${}^7{\rm Be}
(p,\gamma) {}^8{\rm B}$ $S$-factors of $11.2\pm1$~eV-b,
$11.5\pm2.5$~eV-b, and $12.3\pm3$~eV-b at $E_{17}=0.6$, 0.8, and
1.0~MeV, respectively. Using the rather reliably known energy
dependence of the ${}^7{\rm Be} (p,\gamma) {}^8{\rm B}$ $S$-factor
\cite{Kim,Descouvemont}, these values extrapolate to $S(20~{\rm keV})
=12\pm3$~eV-b. This value is about 25\% smaller than the $S$-factor
derived from the same data in Ref.~\cite{Gai}, and it is only 55\%
(62\%) of the $S$-factor adopted in the most recent version of
Bahcall's \cite{Bahcall1} (Turck-Chieze's \cite{Turck}) Standard
Solar Model. We note that such a low value of $S(20~{\rm keV})$
brings the predicted flux of high-energy neutrinos in agreement with
the observation of Kamiokande III \cite{Kamio}. Thus, it is obviously
very important to determine the role the $E2$ Coulomb break-up plays
in the ${}^8{\rm B}+ {}^{208}{\rm Pb} \rightarrow p+{}^7{\rm Be}
+{}^{208}{\rm Pb}$ dissociation process at low energies.

The $S$-factor extracted here from the ${}^8{\rm B}+ {}^{208}{\rm Pb}
\rightarrow p+{}^7{\rm Be} +{}^{208}{\rm Pb}$ data is noticeably
smaller and incompatible (within 2 standard deviations) with the one
recently derived from the various direct measurements of the
${}^7{\rm Be} (p,\gamma) {}^8{\rm B}$ reaction \cite{Johnson}. As it
is important to resolve this apparent difference between the two
methods, a precise direct capture experiment at one energy to pin
down the overall normalization of the direct capture results is
highly desirable. A confirmation of the Coulomb dissociation data and
a verification of its assumed relation to the capture cross section
is also desirable.

In summary, we have shown that the $E2$ component in the ${}^8{\rm
B}+ {}^{208}{\rm Pb} \rightarrow p+{}^7{\rm Be} +{}^{208}{\rm Pb}$
break-up can have a sizeable effect at low energies, in contrast to
the assumption of a previous analysis of ${}^8{\rm B}+ {}^{208}{\rm
Pb} \rightarrow p+{}^7{\rm Be} +{}^{208}{\rm Pb}$ data, which ignored
the $E2$ contributions \cite{Gai}. If our conjecture is confirmed,
the data of Ref.~\cite{Gai} result in a ${}^7{\rm Be} (p,\gamma)
{}^8{\rm B}$ $S$-factor at solar energies of $12\pm3$~eV-b. This
value is noticeably smaller than the $S$-factors obtained in direct
capture measurements \cite{Johnson,Kavanagh,Filippone} and, if
correct, will obviously have important consequences for the
understanding of the solar neutrino puzzle. Our present estimate for
the $E2$ cross section is based on a simple potential model and
clearly calls for an improved treatment. A more reliable microscopic
calculation based on the framework of the multichannel resonating
group model is currently in progress \cite{Csoto}. However, due its
potential importance for the solar neutrino problem, an experimental
determination of the $E2$ contribution is indispensable. This can be
done by measuring the triple-differential cross section $\frac{d^3
\sigma}{d \Omega_R d \Omega_{17} d E_{17}} $, which is sensitive to
the interference of $E1$ and $E2$ components and should show sizeable
effects of the $E2$ break-up amplitudes, even it is somewhat smaller
than estimated in the presently adopted potential model.

\acknowledgements
The authors thank B. W. Filippone and S. E. Koonin for helpful
comments on the manuscript. This work has been supported in part by
the National Science Foundation (Grants No. PHY90-13248 and
PHY91-15574).

\begin{figure}
\caption{$E2/E1$ ratio of virtual photon numbers (upper panel) and of
partial double-differential cross sections (lower panel) for the
${}^8{\rm B}+ {}^{208}{\rm Pb} \rightarrow p+{}^7{\rm Be}
+{}^{208}{\rm Pb}$ break-up process as a function of energy $E_{17}$
and for various Rutherford angles.}
\end{figure}

\begin{figure}
\caption{Comparison of the $E1$ (dotted curve) to the total $E1+E2$
(dashed curve) Coulomb dissociation yield as a function of the
Rutherford angle at three different energies $E_{17}$. The data and
the $E1$ contributions are from Ref.~\protect{\cite{Gai}}. The solid
curve shows the best-fit to the data, including $E1$ and $E2$
contributions, as described in the text.}
\end{figure}

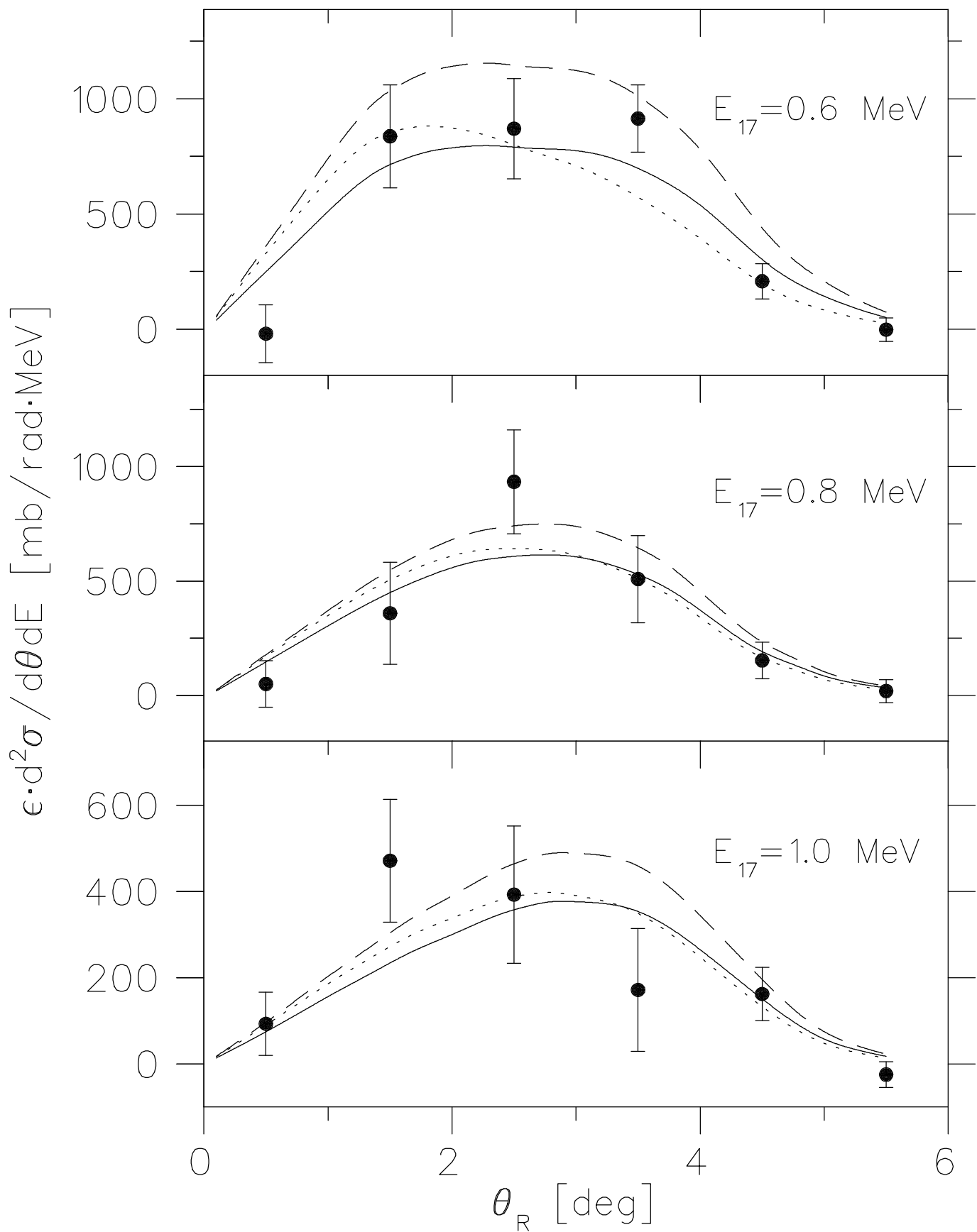
\end{document}